\documentclass[aps,prl,twocolumn,showpacs,superscriptaddress,groupedaddress,longbibliography,showkeys]{revtex4-1}
\usepackage{graphicx}
\usepackage{newtxtext,newtxmath}
\usepackage{enumerate}
\usepackage{bm}
\usepackage{dcolumn}
\usepackage{geometry}
\begin{document}

\title{Electronic and structural properties of
3D Hopf-linked carbon allotrope: Hopfene}
\author{Isao Tomita}
\email{i.tomita@soton.ac.uk}
\affiliation{Sustainable Electronic Technologies,
School of Electronics and Computer Science,
Faculty of Physical Sciences and Engineering,
University of Southampton, SO17 1BJ, UK}
\affiliation{Department of Electrical and Computer Engineering,
National Institute of Technology, Gifu College, Gifu 501-0495,
Japan}
\author{Shinichi Saito}
\affiliation{Sustainable Electronic Technologies,
School of Electronics and Computer Science,
Faculty of Physical Sciences and Engineering,
University of Southampton, SO17 1BJ, UK}
%\date{\today}

\begin{abstract}
Electronic and structural properties of a 3D carbon allotrope
made of Hopf-linked graphenes, which we call a Hopfene
- a type of topological crystal, are examined by semi-empirical
molecular-orbital and density-functional-theoretical methods,
where band-structure analyses reveal very different properties
from those of 2D graphenes.  Furthermore, the analyses give
an interesting finding that, depending on graphene-sheet spacings,
Hopfenes exhibit different band features between primary-type Hopfene
with a finite minimum sheet spacing and secondary type
with its double-sized spacing.  The primary type shows
semi-metallic nature and the secondary type exhibits
semi-metallic or semiconducting nature at different bands
and also has flat bands; these conducting features
can be utilised by Fermi-level control. A device application
of Hopfenes is also provided.
\end{abstract}

%\begin{keywords}
%Band structure, electrical conduction;
%graphene; Hopf-link; topology
%\end{keywords}

\maketitle

\section{Introduction}
The discovery of carbon allotropes \cite{Kroto,Iijima,Nov1}
unveiled their unusual fundamental physical properties via
electrical and spectroscopic analyses \cite{Ando,Ohshima,Ferrari},
albeit using only one type of material - carbon.
Their scientific applications harnessing morphological features
together with electronic states realised there have been powerfully
made in condensed-matter physics; for example, superconductivity
in alkali-metal-doped fullerenes \cite{Super},
Tomonaga-Luttinger liquid in carbon nanotubes \cite{Lutt},
and quantum Hall effect (QHE) \cite{Nov2,FQHE} and
Kosterlitz-Thouless (KT) transition \cite{KT} in graphenes.
Of these, the findings of fullerenes and graphenes,
which imply those outstanding scientific applications,
received Nobel chemistry and physics prizes in 1996 and 2010,
respectively, and also their useful functionalities have
since then spread all over material science and nano-electronics.

%%%%%%%%%%%%%%%%%%%%%%%%% Figure 1 %%%%%%%%%%%%%%%%%%%%%%%%%%%%%
\begin{figure}[htbp]
\begin{center}
\includegraphics[width=80 mm]{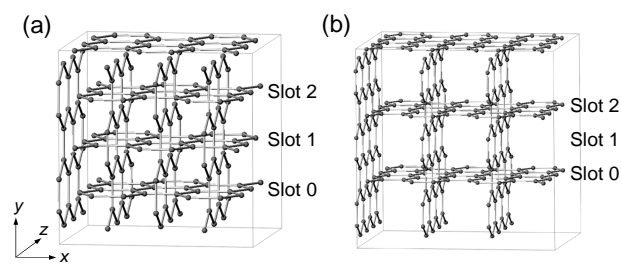}
\end{center}
\caption{(a) (1,1)-Hopfene without any empty slots in both
horizontal and vertical directions. (b) (2,2)-Hopfene
with an empty slot at e.g., slot 1 between
two slots at slot 0 and 2 in the horizontal direction,
and the same arrangements are made in the vertical direction.
The scales for showing (a) and (b) are different, but
the hexagon sizes in (a) and (b) should be the same.}
\label{fig1}
\end{figure}
%%%%%%%%%%%%%%%%%%%%%%%%%%%%%%%%%%%%%%%%%%%%%%%%%%%%%%%%%%%%%%%%%

On the other hand, 'topology' is now one of the important
keywords to deeply understand new kinds of condensed-matter
physics, such as QHE in terms of the Chern number,
KT transition caused by vortex pairs (topological-defect pairs),
and topological insulators and superconductors.
Inspired by those new standpoints,
we have proposed a variety of topological carbon structures
in the form of a molecule, a chain, a chainmail, and
a crystal \cite{S+T_1}.  Among them, such topological crystals
made of graphenes as in Figs.~\ref{fig1}(a)(b)
are of particular interest (where their stability is kept
by insertion of finite-sized vertical graphene layers
into infinitely-large horizontal graphene layers \cite{S+T_2};
to set stress free, the crystal size here is assumed
to be much greater than atomic-structure periods in the crystal).

Despite using graphenes, those topological crystals may exhibit
qualitatively different electrical properties from the graphenes
with semiconducting nature \cite{Nov3} because of different
topologies; however, their precise characteristics are not yet
clear, not even clear whether the proposed crystals are
metallic, semi-metallic, or insulating.  The present paper
clarifies this point by band-structure analyses using
a density-functional-theoretical (DFT) method \cite{DFT0,DFT1,DFT2},
where semi-empirical molecular-orbital (MO) analyses \cite{MOPAC}
are also used for structural analyses.

Structurally-optimized crystals, as shown in
Figs.~\ref{fig1}(a)(b) via the semi-empirical MO method,
have Hopf-links at their intersections.  We thus call these
crystals \emph{Hopfenes} after topologist Heinz Hopf
who studied those links in detail.  The Hopfenes
in Figs.~\ref{fig1}(a)(b) have atomic symmetry P4$_2$bc and
P4$_2$, respectively.

%%%%%%%%%%%%%%%%%%%%%%%%%% Figure 2 %%%%%%%%%%%%%%%%%%%%%%%%%%%%%
\begin{figure}[htbp]
\begin{center}
\includegraphics[width=65 mm]{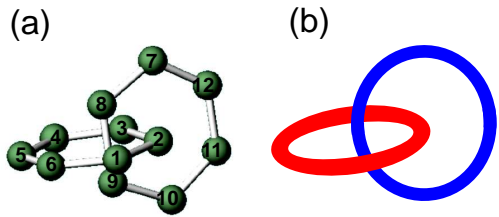}
\end{center}
\caption{(a) Carbon Hopf-link - the building block of Hopfenes,
where the balls represent carbon atoms and the bars depict
$\sigma$-bonds.  The bond length between No.~2 and 3
(and also No.~8 and 9) is extended by 20 \%
(the others by 9 \%) from the original hexagon bond length (1.42 \AA).
A bond-length extension by $>20$ \% is allowed in some hydrocarbons
\cite{Chem}.  When a Hopfene is perfectly formed, the bond between
carbon No.~8 and 9 moves to the centre of the hexagon
with No.~1 - 6. (b) Schematic diagram of a Hopf-link;
in chemistry, it is called a catenane.}
\label{fig2}
\end{figure}
%%%%%%%%%%%%%%%%%%%%%%%%%%%%%%%%%%%%%%%%%%%%%%%%%%%%%%%%%%%%%%%%%

A Hopf-link, as illustrated in Fig.~\ref{fig2}(a), is
the building block of Hopfenes.  Calculated energy of
forming a single Hopf-link is high ($\sim 600$ kcal/mol),
so that the single link will require a high compressive pressure
for its stable existence in nature (thus it may be found
in coke stones at a place deep under the ground).
Embedding Hopf-links in the Hopfene helps the link stability.
In Fig.~\ref{fig2}(b), or a schematic diagram of a link
when including more than six carbon atoms in its ring,
the formation energy is lowered, and the link is realised
by chemical synthesis at 1 atm  \cite{Itami}.
If polymers are used in place of carbons, it is known
that the Hopfene can be achieved \cite{MacGill,Prose1,Prose2}
(but with slightly different atomic symmetry I4/mcm).
The nature of such topological crystals as Hopfenes is interesting,
but their electrical conduction is unknown; the electrical
conduction is a very important property in crystals.
The present paper thus focuses on the study of electrical
conduction via band-structure analyses, which will show
very different features from those of 2D graphenes;
furthermore, a difference in graphene-layer spacing
will provide different properties in electrical conduction.

The paper is organized as follows: \S~\ref{CHS} provides
the details of carbon Hopfene structures; \S~\ref{WBS} gives
analysed results of wavefunctions and band structures
of the Hopfenes and also discusses the obtained results;
and \S~\ref{Cncl} provides a summary.

\section{Carbon Hopfene structures}
\label{CHS}

Hopfenes have some different types,
as seen in Figs.~\ref{fig1}(a)(b),
with respect to graphene-sheet spacings and
sheet phase-shifts.  Figure \ref{fig1}(a) shows that
the slot-1 sheet inserted horizontally (i.e., parallelly
to the $x$-$z$ plane) is set to the exact middle
between slot-0 and slot-2 sheets so that
\emph{horizontal $\sigma$-bond
ladders between zigzag edges intersect
vertical $\sigma$-bond ladders between other
zigzag edges}; this causes a half-lattice-constant shift
(or a phase '$\pi$'-shift) at the slot-1 sheet as long as
the slot-1 sheet is set between slot-0 and slot-2 sheets
(i.e., with no empty slots), where slot-0 and
slot-2 sheets have no phase shift for each other.
This configuration precludes the insertion of other sheets
in parallel to the $x$-$y$ plane.  We call this type
of Hopfene a (1,1)-Hopfene.
On the other hand, Fig.~\ref{fig1}(b) depicts a Hopfene
where the slot-1 sheet is removed (for other parts
in the horizontal direction, similar removal is made);
the same sheet removal in the vertical direction
is also made.  We call this type of Hopfene
a (2,2)-Hopfene, which has no phase-shifted sheets
for parallelly-aligned sheets.

Both (1,1)- and (2,2)-Hopfenes have a tetragonal unit cell
with lattice constants of $a$ and $b \, (= a)$ in the $x$
and $y$ directions and $c \, (\neq a = b)$ in the $z$ direction.
It seems that the period in Fig.~\ref{fig1}(b) in the $x$ and $y$
directions is double as large as that in Fig.~\ref{fig1}(a).
But since the graphene sheet at slot 1 in Fig.~\ref{fig1}(a)
is '$\pi$'-shifted in the $z$ direction, both periods
in the $x$ and $y$ directions in Figs.~\ref{fig1}(a)(b)
are the same.  In Fig.~\ref{fig1}(a), although there remains
the phase '$\pi$'-shift at the slot-1 sheet in the $z$ direction,
the Hopfene periods in the $z$ direction both
in Figs.~\ref{fig1}(a)(b) are actually the same;
the difference is the cross-sectional atomic configuration.

\section{Wavefunction and band-structure analyses}
\label{WBS}

\begin{figure}[htbp]
\vspace*{5 mm}
\begin{center}
\includegraphics[width=75 mm]{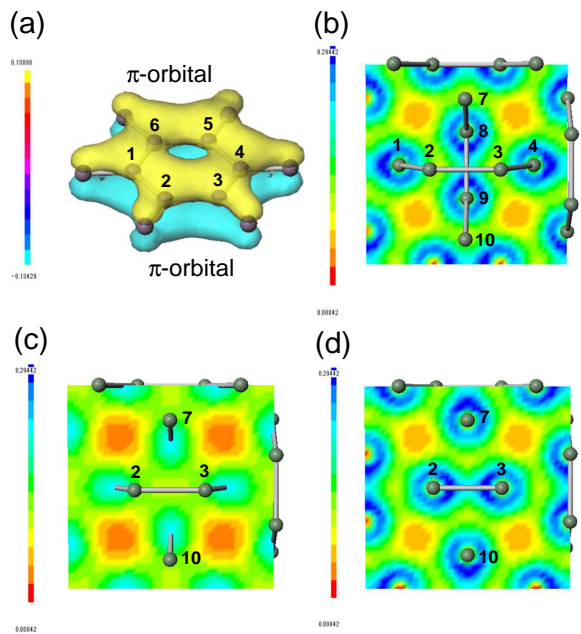}
\end{center}
\caption{(a) $\pi$-orbital wavefunctions
formed on the upper and lower graphene surfaces,
where the vertical axis is in a normalized unit.
(b) Electron probability distribution patterns including
horizontally and vertically extended $\pi$-orbitals,
indicated in yellow (normalized unit)
for the (1,1)-Hopfene, occupying the intermediate
region of the Hopf-link at the cross-section containing
carbon No.~1, 4, 8, 9. The carbon atoms have
localised electrons around them, indicated in blue.
Regions with very small wavefunction amplitudes
(or almost no electron regions) are indicated in orange.
(c) Intermediate cross-section between (b) and (d),
where the extended $\pi$-orbitals almost fully occupy
the region, except the 'orange' holes; localised electrons
around the $\sigma$-bonds are indicated in light blue.
(d) Similar to (b) but at the cross-section containing
carbon No.~2, 3, 7, 10.  For carbon No.~7, 8, 9, 10,
refer to Fig.~\ref{fig2}(a).}
\label{fig3}
\end{figure}

We then analyse electron wavefunctions and band structures
of the Hopfenes by the DFT method.
As is well-known, the band structure of 3D molecular crystals
made up of such molecules as 2D saturated hydrocarbons
with $\sigma$-bonds is not so much different in bandgap
from that of isolated molecules with a HOMO-LUMO energy gap.
But, forming a honeycomb structure (without hydrogens),
i.e., a graphene, changes the situation, producing
an electronic band with a zero gap, where extended $\pi$-orbital
wavefunctions are allowed, as seen in Fig.~\ref{fig3}(a)
on the upper and lower graphene surfaces; this also recovers
electrical conduction (if the Fermi level is set higher
or lower than the \emph{Dirac} point by donor- or
acceptor-doping or by gate-field application
when using a field-effect-transistor structure).

%%%%%%%%%%%%%%%%%%%%%%%%%% Figure 3 %%%%%%%%%%%%%%%%%%%%%%%%%%%%%
%%%%%%%%%%%%%%%%%%%%%%%%%%%%%%%%%%%%%%%%%%%%%%%%%%%%%%%%%%%%%%%%%

In the (1,1)-Hopfene, extended $\pi$-orbitals, obtained
by numerically solving the Kohn-Sham equation, go
in the vertical direction as well as in the horizontal direction,
as displayed in yellow in Figs.~\ref{fig3}(b)-(d); these figures
include other Kohn-Sham-wavefunction contributions painted
in blue and light blue.  These blue and light blue ones are
localised electrons around the carbon atoms and the $\sigma$-bonds,
respectively.  From Figs.~\ref{fig3}(b)(c), we can see that
rings of the Hopf-link in the extended $\pi$-orbitals are
standing well, supported by repulsive force from
the localised electrons in blue and light blue.
As seen in Fig.~\ref{fig3}(c), regions with very small
wavefunction amplitudes (i.e., regions with almost no electrons)
are shown in orange.  It is not shown, but the (2,2)-Hopfene
has very similar electron distributions, while having
much bigger 'holes' at orange regions.

%%%%%%%%%%%%%%%%%%%%%%%%%% Figure 4 %%%%%%%%%%%%%%%%%%%%%%%%%%%%%
\begin{figure}[htbp]
\begin{center}
\includegraphics[width=80 mm]{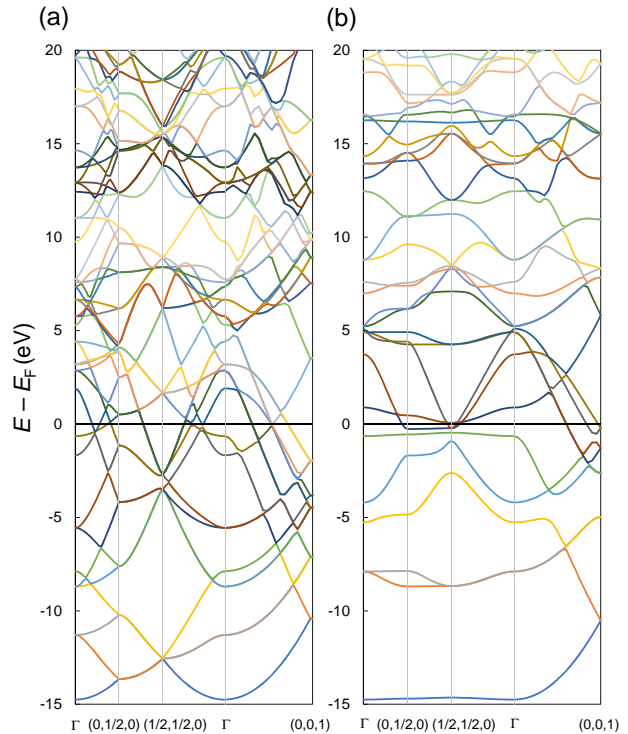}
\end{center}
\caption{Band structures of (a) (1,1)-Hopfene and
(b) (2,2)-Hopfene, where the lattice constants are
$a = b = 5.1$ \AA \,
and $c = 2.6$ \AA.  Here, the electron energy $E$
in the vertical axis is measured
from the Fermi level $E_{\rm F}$.}
\label{fig4}
\end{figure}
%%%%%%%%%%%%%%%%%%%%%%%%%%%%%%%%%%%%%%%%%%%%%%%%%%%%%%%%%%%%%%%%%

The obtained band structures (Kohn-Sham-eigenvalue
distributions) of the (1,1)- and (2,2)-Hopfenes are presented
in Figs.~\ref{fig4}(a)(b), respectively; these show
rather dissimilar structures, despite the difference
just in the 'hole' size.  We also observed
several flat bands in Fig.~\ref{fig4}(b).  So far
the mechanism of this flat-band appearance has not yet
been well understood; but we have simultaneously
observed a great decrease in the number of
dispersion branches with branch rearrangement
(with increasing degeneracy).  We thus infer that
this flat-band appearance is intimately related to
this increased degeneracy, which results from
the higher crystal symmetry of the (2,2)-Hopfene
\emph{with no phase-shifted sheets} than that of
the (1,1)-Hopfene \emph{with a large number of
phase-shifted sheets}.  The presence of flat bands
is interesting, and actually very useful
in solid-state physics; for example, this makes it
possible to study the ground-state property of
a many-body carrier system in a nonperturbative manner,
and the Hopfene can be a type of
itinerant carrier system that provides such a possibility.

As for details of the band structure of the (1,1)-Hopfene
in Fig.~\ref{fig4}(a), we can see that
the band is clearly \emph{semi-metallic} because
it has partially filled conduction and valence bands
(valleys), which are caused by the graphene-intertwined
structure and thus clearly differ from the band of
2D graphenes.  In Fig.~\ref{fig4}(b) for the (2,2)-Hopfene,
we can see semi-metallic or semiconducting parts
at different bands and also see several flat bands;
those bands can be utilised by Fermi-level shifting
with, e.g., doping. (But too much doping should be
avoided to prevent a large band-structure deformation.)
Here, since the flat bands have high density of states
(DOS) with carriers having very large effective mass,
if the Fermi level is set to a flat band,
the electron-interaction term is dominant in the competition
between electron kinetic-energy and interaction terms
in an $e$-$e$-interaction Hamiltonian, such as
the Hubbard Hamiltonian, which could cause
ferromagnetism \cite{FB1,FB2,FB3,FB4}.
Research on Hopfenes is very interesting, and we expect that
more interesting properties will come from it,
including BCS-BES crossover and superconductivity-insulator
transition \cite{S+T_3}.

On the application side of Hopfenes, since chemical sensors
with a graphene, for example, currently have a low adsorption
rate of gas molecules on the graphene, if we replace
the graphene with a Hopfene with some electrical conduction
and \emph{use Hopfene's holes (nanopores) to effectively capture
the gas molecules}, the sensor will obtain a high adsorption rate,
thus enhancing its sensitivity, where sensing is made via
a change in the electrical conduction when molecule adsorption
occurs.

\section{Conclusions}
\label{Cncl}
We have proposed Hopf-linked carbon allotropes, named Hopfenes,
with the help of structural analyses via a semi-empirical MO method
(where Hopfenes are assumed to be much larger in size than
their lattice periods and supported in large graphenes).
We then have analysed the electron wavefunctions of a Hopfene
by DFT analyses, which show that its Hopf-links in extended
$\pi$-orbitals are standing well (n.b., polymer Hopfenes
in the same situation are known to exist experimentally).
The DFT band-structure analyses have exhibited that unlike
2D graphenes with semiconducting nature with a zero gap,
(1,1)-Hopfenes are semi-metallic and that (2,2)-Hopfenes
are semi-metallic or semiconducting at different bands
and have several flat bands; these can be utilised
by Fermi-level control.  The proposed carbon Hopfenes
have not yet been realised, but if they are produced or found,
they can be used for interesting electrical-conduction
and magnetism research and also for sensor applications.

\section*{Acknowledgements}
We would like to thank Prof. H. Mizuta,
Dr. M. Muruganathan, Prof. Y. Oshima, Prof. S. Matsui,
Prof. S. Ogawa, Prof. S. Kurihara, and Prof. H. N. Rutt
for stimulating discussions.  S.S. would also like
to thank the Center for Single Nanoscale Innovative
Devices of JAIST. The data from the paper can be
obtained from the University of Southampton ePrint
research repository:
https://doi.org/10.5258/SOTON/D0919

\section*{Disclosure statement}
No potential conflict of interest was reported by the authors.

\section*{Funding}
This research is support by EPSRC Manufacturing Fellowship
(EP/M008975/1).

\end{document}